

\input phyzzx.tex
\overfullrule=0pt

\def\sun{{\rm SU}(N)}
\def\un{{\rm U}(N)}
\def\spn{{\rm Sp}(N)}
\def\son{{\rm SO}(N)}
\def\Sbar{ {\overline S}}
\def\cS{{\cal S}}
\def\Eu{{\cal E} }
\def\sinS{  \sigma \in S_r }
\def\RinY{  R \in Y_r}
\def\col{k}
\def\e{  {\rm e} }

\def\Wk { W_\kappa }
\def\nW { {\widehat W} }
\def\nWk{ \nW_\kappa }
\def\ie {{\it i.e.}}
\def\ts{\textstyle}
\def\half{{\ts{1\over 2}}}
\def\threehalf{{\ts{3\over 2}}}
\def\largeNlimit{\mathrel{\mathop{\longrightarrow}\limits_{N \to \infty}}}
\def\PLB{ \sl Phys. Lett.         \bf B}
\def\NPB{ \sl Nucl. Phys.         \bf B}
\def\PRD{ \sl Phys. Rev.          \bf D}

\def\MPLA{ \sl Mod. Phys. Lett. \bf  A}
\def\IJMPA{ \sl Int. Jour. Mod. Phys. \bf  A}
\def\CMP{ \sl Commun. Math. Phys. \bf  }
%
\newbox\bigstrutbox
\setbox\bigstrutbox=\hbox{\vrule height12pt depth5.5pt width0pt}
\def\tablerule{\noalign{\hrule}}
\def\thtablerule{\noalign{\hrule height1pt}}
\def\bigstrut{\relax\ifmmode\copy\bigstrutbox\else\unhcopy\bigstrutbox\fi}
%
\def\one{  {\vcenter  {\vbox
              {\hrule height.4pt
               \hbox {\vrule width.4pt  height3pt
                      \kern3pt
                      \vrule width.4pt  height3pt }
               \hrule height.4pt}
                         }
                   }
           }
\def\barone{ \overline{\one}}
\def\two{  {\vcenter  {\vbox
              {\hrule height.4pt
               \hbox {\vrule width.4pt  height3pt
                      \kern3pt
                      \vrule width.4pt  height3pt
                      \kern3pt
                      \vrule width.4pt height3pt}
               \hrule height.4pt}
                         }
              }
           }
\def\oneone{ {\vcenter  {\vbox
              {\hrule height.4pt
               \hbox {\vrule width.4pt  height3pt
                      \kern3pt
                      \vrule width.4pt  height3pt }
               \hrule height.4pt
               \hbox {\vrule width.4pt  height3pt
                      \kern3pt
                      \vrule width.4pt  height3pt }
               \hrule height.4pt}
                         }
              }
           }

\baselineskip=14pt
\line{hep-th/9503020 \hfil BERC-94/102}
\line{March, 1995 \hfil BOW-PH-104}
\line{\hfil BRX-TH-366}
\bigskip
\centerline{\seventeenrm Large N Universality of the }
\centerline{\seventeenrm Two-Dimensional Yang-Mills String}
\bigskip
\centerline{\twelvecp Michael Crescimanno
\footnote\${Research supported in part by the DOE under
grant DE-FG02-92ER40706}
\footnote\dag{crescimanno@berea.edu}}
\centerline{Department of Physics}
\centerline{Berea College}
\centerline{Berea, KY  40404}
\medskip
\centerline{\twelvecp Stephen G. Naculich
\footnote\star{naculich@polar.bowdoin.edu}}
\centerline{Department of Physics}
\centerline{Bowdoin College}
\centerline{Brunswick, ME  04011}
\smallskip
\centerline{and}
\smallskip
\centerline{\twelvecp Howard J. Schnitzer$^\$ $
\footnote\ddag{schnitzer@binah.cc.brandeis.edu} }
\centerline{Department of Physics}
\centerline{Brandeis University}
\centerline{Waltham, MA  02254}
\baselineskip=\normalbaselineskip
\abstract{
We exhibit the gauge-group independence (``universality'')
of all normalized non-intersecting Wilson loop expectation values
in the large $N$ limit of two-dimen\-sion\-al Yang-Mills theory.
This universality is most easily understood
via the string theory reformulation of these gauge theories.
By constructing an isomorphism between the string maps
contributing to normalized Wilson loop
expectation values in the different theories,
we prove the large $N$ universality of these observables
on any surface.
The string calculation of the Wilson loop expectation value on the sphere
also leads to an indication of the large $N$ phase transition
separating strong-  and weak-coupling phases.}
\endpage

%
\chapter{Introduction}

\REF\rMR{ A. Migdal, \sl Sov. Phys. JETP \bf 42 \rm (1975) 413.}
\REF\rLoop{ N. Bralic, \PRD 22 \rm (1980)  3090;\nextline
     V. Kazakov and I. Kostov, \NPB 176 \rm (1980) 199.}
\REF\rExact{ B. Rusakov, \MPLA 5 \rm (1990) 693; \nextline
 D. Fine, \CMP 134 \rm (1990) 273; {\bf 140} (1991) 321; \nextline
 E. Witten, \CMP 141 \rm (1991) 153; \nextline
 M. Blau and G. Thompson, \IJMPA 7 \rm (1992) 3781.}
\REF\rEarly{ G. 't Hooft, \NPB 75 \rm (1974) 461; \nextline
 W. Bardeen, I. Bars, A. Hanson, and R. Peccei,
                 \PRD 13 \rm (1976) 2364.}
\REF\rGTM{ D. Gross, \NPB 400 \rm (1993) 161; \nextline
 J. Minahan, \PRD 47 \rm (1993) 3430; \nextline
 D. Gross and W. Taylor, \NPB 400 \rm (1993) 181.}
\REF\rTwist{ D. Gross and W. Taylor, \NPB 403 \rm (1993) 395.}
\REF\rNRS{S. Naculich, H. Riggs, and H. Schnitzer,
 \MPLA 8 \rm  (1993) 2223; \PLB 466 \rm (1993) 466.}
\REF\rRam{S. Ramgoolam, \NPB 418 \rm (1994) 30.}
\REF\rNRSwilson{S. Naculich, H. Riggs, and H. Schnitzer,
 hep-th/9406100, \IJMPA \rm, in press;\nextline
 S. Naculich and H. Riggs, hep-th/9411143, \PRD \rm, in press.}
\REF\rEuler{S. Ramgoolam, hep-th/9412110.}

Pure Yang-Mills theory on an arbitrary
(Euclidean) two-manifold  is an
exactly solvable quantum field theory [\rMR--\rExact].
It has been shown that these theories
are string theories [\rEarly--\rEuler],
although generalizing this discovery to
gauge theory in higher dimensions has proven elusive.

Wilson loop vacuum expectation values (VEVs)
form a complete basis of gauge-invariant observables
of the pure gauge theory.
An important class of these observables,
the VEVs of non-intersecting Wilson loops,
can be written as exact expressions that depend on
the representation theory of the gauge group $G$ [\rExact].
For the Lie groups $G = \sun$, $\un$, $\son$, or $\spn$,
one may expand the Wilson loop observables
in $1/N$ about $N = \infty$;
the large $N$, strong-coupling limit
is the natural one for making explicit the connection
between gauge theory and string theory.

\REF\rCS{M. Crescimanno and H. Schnitzer, hep-th/9501099.}
\REF\rDK{M. Douglas and V. Kazakov, \PLB 319 \rm (1993) 219.}
When one computes simple Wilson loops,
one discovers that the normalized VEV is independent,
in the large $N$ limit,
of the choice of gauge group $G$
(e.g., see eqs. (2.3), (2.4), and (2.5) below).
This suggests that there is but one
theory at large $N$ in two dimensions.
At first sight,
this conclusion may strike one as either incorrect,
or else trivial.
We show below that it is correct;
here we describe why it is non-trivial.
First, in two-dimensional Yang-Mills theory at weak coupling,
the form of the gauge field propagators
combined with the Feynman diagrams
in leading order in $1/N$
suggest that the computation of
any {\it perturbative} gauge-invariant quantity
must be the same for all groups at large $N$.
What is claimed here and in ref.~[\rCS], however,
is the universality of observables {\it for all coupling}.
The third order phase transition separating the
strong- and weak-coupling regions of the theory on a
two-sphere [\rDK] complicates extending
the perturbative equivalence to equivalence at all coupling.
Second,  the exact gauge theory expressions for Wilson loop VEVs
appear to depend on many details of the representation theory
of the group $G$,
and this large $N$ equivalence
requires highly non-trivial cancellations
of {\it sub-leading} (in $1/N$) terms
coming from the dimensions and quadratic Casimirs of
representations.
{}From the gauge theory point-of-view,
large $N$ universality seems miraculous.

Large $N$ gauge-group independence was first noticed in ref.~[\rCS],
where the large $N$ saddle point
on the sphere (and projective plane) was
computed for various gauge groups.
There it was shown that the saddle point is universal,
and the computation of the partition function
and non-intersecting Wilson loop
for the fundamental representation
led to the same result for all groups $G$ at large $N$.
{}From the point of view of
this matrix-model-like, large $N$ technique,
universality
is the result of one's ability to rewrite the expressions
for the free energy and Wilson loop on the sphere
in a universal form.
This does not reveal any deeper reason for the universality.

In this paper, we demonstrate that
the string theory reformulation
of two-dimensional Yang-Mills theory
provides the natural framework for understanding
large $N$ universality.
We show that the large $N$ equivalence of Wilson loop VEVs
follows from an isomorphism between string maps
in the various theories.
We prove an exact
(to all orders in the double expansion
in the areas and the exponential of the areas)
equality of normalized VEVs of non-intersecting Wilson loops
in an arbitrary representation on an arbitrary surface
for all the classical Lie groups
to leading order in $1/N$.
This is a significant demonstration of the
utility of the string theory reformulation of
two-dimensional Yang-Mills theory.

%
\chapter{Wilson loop universality}

Consider a homologically trivial Wilson loop
that divides an orientable surface $\cS$
of genus $h$ and (dimensionless) area $A$
into a disk $D$ with area $A_1$
and a surface $\cS \backslash D$ with area $A_2$.
The expectation value of the trace of this Wilson loop
in some representation $R$ of the gauge group $G$
is [\rMR--\rExact]
$$
W_R
= \sum_{R_1, R_2} N_G(R_1, R; R_2) (\dim R_1) (\dim R_2)^{1-2h}
\exp \left[ -~ { A_1 C_2(R_1) + A_2 C_2(R_2)  \over 2Nf } \right]
\eqn\eWR
$$
where $C_2(R)$ and $\dim R$ denote
the quadratic Casimir and dimension of the representation $R$,
$N_G(R_1, R; R_2)$ are tensor product multiplicities of the group $G$,
and $f$ is $\half$ for $\spn$ and $1$ otherwise.
In the string theories corresponding to these Yang-Mills theories,
the natural observables are not $W_R$
but rather the linear combinations [\rTwist, \rRam, \rNRSwilson]
$$
\Wk = \sum_{\RinY} \chi_R (\kappa) W_R
\eqn\eWkap
$$
where $\kappa$ denotes a conjugacy class of the symmetric group $S_r$,
$Y_r$ denotes the set of Young diagrams
with $r$ cells,
and $\chi_R (\kappa) $ is the character of the conjugacy class $\kappa$
in the representation $R$ of $S_r$.
The $\Wk$ basis,
not the irreducible representation basis $W_R$,
is the appropriate basis
in which to compare observables between different gauge theories.

\REF\rBoul{D. Boulatov, \MPLA 9 \rm (1994) 365.}
We claim that the large $N$ limit of the normalized VEV
$\nWk = \Wk /Z $
for non-intersecting Wilson loops (on any surface)
is independent of the gauge group.
First, we give several examples.
Consider the homologically trivial Wilson loop on the sphere
with $\kappa = (1)$,
a single cycle of length one.
Irrespective of the gauge group,
the first few terms of its normalized VEV are
\foot{Expressions of this type were first given in ref.~[\rLoop].}
\foot{This corrects an error in ref.~[\rBoul]. See appendix of
ref.~[\rCS].}
$$
\eqalign{
\nW_{(1)}  = {W_\one \over Z}
& \largeNlimit
 N \bigg[ \e^{-\half A_1 }
+  \left(\half A_1^2 - A_1 - 1 \right) \e^{-A_1 - \half A_2} \cr
& +  \big(\half A_1^4 + {\ts{2\over 3}} A_1^3 A_2
+ {\ts{1\over 4}} A_1^2 A_2^2 + {\ts{8\over 3}} A_1^3
- {\ts{5\over 2}} A_1^2 A_2 - \half A_1 A_2^2 + \threehalf A_1^2 \cr
& + A_1 A_2 - \half A_2^2 + A_1 + A_2 + 1 \big)
\e^{-\threehalf A_1 - A_2} + \cdots \bigg]  + ( A_1 \leftrightarrow A_2 )
}
\eqn\eWone
$$
Likewise, for $\kappa = (2)$,
a single cycle of length two,
$$
\eqalign{
\nW_{(2)} & = { W_\two - W_\oneone \over Z}  \cr
& \largeNlimit N \bigg[ \left(1 - A_1 \right) \e^{- A_1 }
+  \left(- {\ts{4\over 3}} A_1^3 + 4 A_1^2 \right)
\e^{ - \threehalf  A_1 - \half A_2} + \cdots \bigg]
 + ( A_1 \leftrightarrow A_2 )
\cr}
\eqn\eWtwo
$$
The Wilson loop with $\kappa = (1^2)$ has VEV
$$
\eqalign{
\nW_{(1^2)}  = { W_\two + W_\oneone \over Z}
& \largeNlimit
  N^2  \bigg[ \e^{-A_1 }   + \e^{-\half A_1 -\half A_2}
  +  \left( A_1^2 - 2 A_1 - 2 \right) \e^{- \threehalf A_1 - \half A_2} \cr
& +  \left( \half A_1^2 + \half A_2^2 - A_1 - A_2 - 2\right)
      \e^{- A_1 - A_2} + \cdots \bigg]  + ( A_1 \leftrightarrow A_2 )
\cr}
\eqn\eWonesq
$$
Observe that
$ \nW_{(1^2)} = \left[ \nW_{(1)} \right]^2 $
in the large $N$ limit;
we will see that
this has a natural explanation in string theory.
This large $N$ factorization
is also expected from the existence of the
large $N$ saddle point (closely related to the master field)
but is only valid for representations
whose diagrams
have a finite number of cells.
At the end of section 3,
we give examples of the universal large $N$ limits
of Wilson loops (both homologically trivial and non-trivial)
on surfaces other than the sphere.

\REF\rRusakov{B. Rusakov, \PLB 303 \rm (1993) 95.}
\REF\rMP{J. Minahan and A. Polychronakos, \NPB 422 \rm (1994) 172.}
\REF\rDaul{J. Daul and V. Kazakov, \PLB 355 \rm (1994) 371; \nextline
B. Rusakov, \PLB 329  \rm (1994) 338.}
\REF\rCT{ M. Crescimanno and W. Taylor, hep-th/9408115.}
The large $N$ universality of the
normalized Wilson loop VEVs exhibited above
begs a deeper explanation.
The saddle point techniques [\rDK--\rCT]
used in ref.~[\rCS]
give an analytical view of this universality for simple cases,
but this approach somewhat obscures the physical grounds for the
equivalence.
It is more enlightening to describe universality
as an equivalence between string theories,
to which we now turn.

%
\chapter{String theory explanation}

To make the translation from gauge theory to string theory,
recall that the features of the string maps are determined
by the form of the quadratic Casimir and the dimension of the
representations in eq.~\eWR.
For the classical Lie groups,
the quadratic Casimir for tensor representations is
$$
C_2 (R)
 = fN \left[ r  + {T(R) \over N} - U(r)  \right],
\eqn\eCas
$$
where
$$
\eqalign{
f
& = \cases{ 1   & for $\sun$ , $\un$, and $\son$, \cr
            {1 \over 2} & for $\spn$, \cr} \cr
T(R)
& = \sum_{i=1}^{{\rm rank~}G} \, \ell_i (\ell_i + 1 - 2i)
     = \sum^{\col_1}_{i=1} \, \ell^2_i - \sum_{j=1}^{\ell_1} \col^2_j \cr
U(r)
& = \cases{r^2/N^2 &  for $\sun$, \cr
                 0 &  for $\un$,  \cr
               r/N &  for $\son$, \cr
             - r/N &  for $\spn$, \cr}
}
\eqn\eGroup
$$
with $\ell_i$ ($\col_j$) denoting the row (column) lengths
of the Young diagram corresponding to the representation $R$,
and $r=\sum_{i=1}^{\col_1} \ell_i$ is the number of cells in the
Young diagram.
The dimension of a representation $R$ of $\sun$
whose Young diagram has a finite number of cells $r$
in the limit $N \to \infty$ can be written  [\rTwist]
$$
\dim R = {N^r \over r!} \chi_R (\Omega_r)
\eqn\eDimR
$$
where
$$
\Omega_r = \sum_{\sinS}  N^{c(\sigma)-r} \sigma \qquad {\rm for~SU}(N)
\eqn\eOmegaSU
$$
with $c(\sigma)$ the number of cycles in the permutation $\sigma$.
The dimension of a ``composite'' representation [\rGTM]
$R \Sbar $ of $\sun$,
where $R$ and $S$ each have Young diagrams
with a finite number of cells,
almost factorizes
$$
\dim (R \Sbar) = \dim (R) \dim (\Sbar)
\left[1 + O \left(1 \over N^2\right)\right].
\eqn\eComposite
$$
The dimensions of tensor representations of $\son$ and $\spn$
are also given by eq.~\eDimR,
but the operator $\Omega_r$ has sub-leading terms [\rNRS--\rNRSwilson]
$$
\Omega_r  = \sum_{\sinS} N^{c(\sigma)-r} \left[ 1
+ O\left(1\over N\right) \right] \sigma
\qquad {\rm for~SO}(N){\rm~and~Sp}(N).
\eqn\eOmegaSO
$$
In the string theory,
the first term in the quadratic Casimir \eCas~weights each map
with the exponential of the worldsheet area,
the Nambu-Goto action.
The second term corresponds to the presence
of an arbitrary number of simple branch points [\rGTM].
Both these terms are universal among the different gauge groups.
The third term in the Casimir corresponds to other
features [\rGTM, \rNRS] of the string maps
(infinitesimal handles, cross-caps,
and orientation-preserving tubes [OPT])
which differ among the gauge groups.
In addition, orientation-reversing tubes (ORT)
arise from sub-leading terms in the quadratic
Casimir of composite representations of $\sun$.
The dim $R$ term also gives rise to
various features of the string maps.
The $\Omega_r$ operator \eOmegaSU~ is called a twist-point operator,
each of its terms permuting the
sheets in the manner of a  multiple branch point [\rTwist];
its leading term is universal among the gauge groups.
The sub-leading terms in ~\eComposite~may be
attributed to the presence of infinitesimal ORTs
connecting cycles of equal length [\rTwist];
we call these ``twist-point tubes.''
The sub-leading terms in ~\eOmegaSO~also
give rise to twist-point tubes,
and to infinitesimal cross-caps on cycles of even length [\rRam].

\bigskip
\hbox{
\hskip-.125truein
\vbox{\offinterlineskip
{\hrule height1pt}
\halign to436pt{\strut#&\vrule#\tabskip=1em plus2em&\quad#\hfil&\vrule#&
\hfil#&\vrule#&\hfil#&\vrule#\tabskip=0pt\cr
\bigstrut&& GAUGE GROUP && $\sun$ && $\son$/$\spn$ & \cr\thtablerule
\bigstrut&& Nambu-Goto Action && $\exp(-rA/2)$ && $\exp(-rA/2)$
&\cr\thtablerule
\bigstrut&&\multispan{5} AREA-DEPENDENT FEATURES   \hfil&\cr\thtablerule
\bigstrut&&~~~Simple Branch Points && $-A/N$ && $-A/N$ &\cr\tablerule
\bigstrut&&~~~Orientation Preserving Tubes (OPT) &&$~A/N^2$ &&
&\cr\tablerule
\bigstrut&&~~~Orientation Reversing Tubes (ORT) &&$-A/N^2$ &&
&\cr\tablerule
\bigstrut&&~~~Handles                          &&$~A/2N^2$&& &\cr\tablerule

\bigstrut&&~~~Crosscaps && && $\pm A/2N$ &\cr\thtablerule
\bigstrut&&\multispan{5} AREA-INDEPENDENT FEATURES   \hfil&\cr\thtablerule
\bigstrut&&~~~Multiple Branch (Twist) Points
&& $1/N^{r-c(\sigma)}$ && $1/N^{r-c(\sigma)}$ &\cr\tablerule
\bigstrut&&~~~ORTs between equal length cycles && $-1/N^2$ &&
&\cr\tablerule
\bigstrut&&~~~Tubes between equal length cycles &&&& $-1/N^2$
&\cr\tablerule
\bigstrut&&~~~Crosscaps on even length cycles   &&&& $\mp 1/N$
&\cr\thtablerule
\bigstrut&\multispan{7} \hfil TABLE I: Weights assigned to string map
features
\hfil\cr
}}\hfill}
\medskip

The weights assigned to each of
the string features discussed above
are summarized in Table I.
\footnote{*}{The string version of the $U(N)$ gauge theory
is very similar to that of $SU(N)$,
differing only in the absence of area-dependent tubes and handles.}
Those features that may appear anywhere on the target space
(such as simple branch points)
are weighted with a factor of area,
so we call them ``area-dependent.''
Features that are immobile
(such as twist points)
we call ``area-independent.''

Another difference between the gauge groups
is that the worldsheets of the $\sun$ and $\un$ string theories
have a definite orientation.
Thus, the map from a worldsheet to an oriented target space $\cS$
is at any point either orientation-preserving or orientation-reversing
[\rGTM].
A given worldsheet can have both orientation-preserving
and orientation-reversing components;
one thus refers to the chiralities
of the various components of the worldsheet.
On the other hand,
worldsheets of the $\son$ and $\spn$ string theories
need not be orientable [\rNRS].

In summary, then,
the string theories corresponding to
$\sun$, $\un$, $\son$, and $\spn$ Yang-Mills theory
have in common
the same worldsheet branch- and twist-point structures.
They differ in terms of possible insertions of
infinitesimal handles, tubes, and crosscaps
(both area-dependent and -independent),
and also in that $\sun$ and $\un$ worldsheets are oriented
whereas $\son$ and $\spn$ worldsheets are not.
To understand the large $N$ universality
of Wilson loops,
we must explain how the differences between the theories
(as shown in Table I)
can lead to the same values for VEVs.

The unnormalized VEV $\Wk$ \eWkap~has
a natural stringy interpretation [\rTwist,\rRam--\rEuler].
Each term in the $1/N$ expansion of $\Wk$ corresponds to
a map from a worldsheet with boundary to $\cS$,
where the worldsheet boundary is mapped to the Wilson loop.
If the worldsheet has Euler characteristic $\Eu$,
then its contribution to $\Wk$ is proportional to $N^\Eu$.
The conjugacy class $\kappa$ determines the
boundary conditions on the maps.
If the conjugacy class $\kappa$ has $c$ cycles,
with cycle lengths $n_i$ ($i=1, \ldots, c$),
then the worldsheet has $c$ boundary components $B_i$,
and each of the curves $B_i$ is mapped
with degree $n_i$ to the Wilson loop
(\ie, traversing the curve $B_i$ once,
one circles the Wilson loop $n_i$ times).
Maps that contribute to $\Wk$
can be from surfaces with arbitrarily large Euler characteristic,
and therefore contribute with arbitrarily large power of $N$.
However, these maps correspond to disconnected worldsheets.
Normalizing the Wilson loop expectation value
by dividing by the Yang-Mills partition function,
$$
\nWk = \Wk / Z
\eqn\enWkap
$$
removes the disconnected ``vacuum graphs,''
and results in a quantity with a
large $N$ limit that we will show is universal.

Having described the possible
connected string maps that contribute to the
normalized Wilson loop VEV,
we now analyze those
which dominate in leading order in $1/N$.
We see immediately that
some of the features discussed above are irrelevant
in leading order.
Any $\sun$ worldsheet with an infinitesimal handle
will be suppressed by $1/N^2$
relative to the same worldsheet with the handle removed,
and so is negligible.
Likewise, any $\son$ or $\spn$ worldsheet
with an infinitesimal crosscap
is down by $1/N$
relative to the surface without it.
(Since the weights
of $\son$ and $\spn$ maps differ
only in the sign of the crosscaps,
it immediately follows that
$\son$ and $\spn$ Wilson loop VEVs
are equal to leading order in $1/N$.)
Also, in general, $\son$ and $\spn$ Wilson loops
can have contributions from nonorientable worldsheets,
but such worldsheets
are always suppressed by at least a factor of $1/N$
relative to a contributing orientable worldsheet.

Since infinitesimal tubes
are associated with a factor of $1/N^2$,
one might suppose that they
would also be irrelevant to leading order in $1/N$.
This is not the case.
Any infinitesimal tube
whose removal leaves a map that contributes
to the normalized Wilson loop VEV
can indeed be ignored.
It is possible, however, that the removal of a tube
could result in a surface with a disconnected vacuum part;
such a surface does not contribute
to the normalized VEV.
If the contribution of this disconnected worldsheet
is of higher order in $N$ than the connected worldsheets,
then the insertion of a tube,
which reduces the $N$ dependence,
can result in a surface
that contributes to the {\it same} order
as the other connected worldsheets.
Thus, to leading order in $1/N$,
we may have to include worldsheets with infinitesimal tubes,
but only if the removal of any tube
results in a disconnected vacuum piece.
Therefore, one concludes that the tube structure of the worldsheets
that contribute to leading order in $1/N$ to the VEV
must be {\it tree-like}, \ie, not have any closed loops.

Clearly,
the set of worldsheet maps with only simple branch points
and twist points
will be equivalent in all the string theories,
irrespective of the gauge group.
It is far from obvious, however, that maps
that include infinitesimal tubes will be identical,
since the tube structures differ between the groups.
We will now show that, in fact,
the set of worldsheet maps is identical for all the groups.
We do this by constructing an isomorphism between these maps:
for each worldsheet of $\son/\spn$ that contributes
to leading order in $1/N$,
we will construct the corresponding worldsheet of $\sun$,
and vice versa.

Let us begin with the normalized VEV $\nW_{(n)}$
of a non-intersecting Wilson loop on the sphere,
where $(n)$ denotes the conjugacy class
with a single cycle of length $n$.
Consider the most general map
that contributes to leading order in $1/N$
to this Wilson loop in $\son$ (or $\spn$) string theory.
The worldsheet consists of a set of connected components
$C_0$, $C_1$, $C_2, \cdots, C_m$,
where $C_0$ has the topology of a disk
and $C_1$, $C_2, \cdots, C_m$ are topologically spheres.
If any of the components had a higher genus,
the Euler characteristic of the worldsheet would be decreased
and the map would be sub-leading in the $1/N$ expansion.
The component $C_0$ is mapped to the target space sphere $\cS$,
and its boundary is mapped to the Wilson loop,
wrapping around the loop $n$ times.
For $n > 1$, there will necessarily be branch points
in the map $C_0 \to \cS$.
Each of the other components
$C_1$, $C_2, \cdots, C_m$
is a covering of the sphere $\cS$ with arbitrary degree $n_i$
and branching number $ 2 n_i - 2 $.
The branch points on $C_0, \cdots, C_m$
can be either simple branch points
associated with $T(R)$ in the Casimir,
or multiple branch points associated with
one of the twist points on $\cS$.
While the branching structure of any component $C_i$
may be very complicated,
all of the groups have exactly
the same set of branched covers
with exactly the same weights.
The different components $C_i$
are connected to each other
in a tree-like structure by twist-point tubes
(the only kind of tubes in $\son$).
In particular, any cycle of sheets at a twist point
can be connected to an equal length cycle of sheets
on another component.
Since there can be no closed loops in the structure,
exactly $m$ tubes connect the $m+1$ components.
The Euler characteristic of the connected surface is thus
$ 1 + 2m - 2m = 1$,
showing it to have the topology of a disk.
Thus, the Wilson loop expectation value $ \nW_{(n)} $
is proportional to $N$ in leading order.

We now describe the $\sun$ map
that corresponds to the one just described for $\son/\spn$.
It has of course the same worldsheet,
but with a chirality assigned to each of the components $C_i$.
The conjugacy class $(n)$ of the Wilson loop
is a linear combination of the products
of either the fundamental representation $\one$ of $\sun$
or of its conjugate $\barone$ ;
this determines the chirality of the Wilson loop itself.
The boundary of $C_0$ is mapped to the Wilson loop,
and the region of $C_0$ bordering the boundary
will be mapped either to one side of the Wilson loop or to the other.
The chirality of $C_0$ is fixed by the chirality of the Wilson loop,
and depends on which side of the Wilson loop it approaches.
What are the chiralities of the other components $C_i$?
In $\sun$,
twist points tubes only connect sheets of opposite chirality,
so any component $C_i$ to which $C_0$ is {\it directly} connected
must have chirality opposite that of $C_0$.
Any components connected directly to these components have
the same chirality as $C_0$, and so forth.
Thus,
the chirality of each of the components of the $\sun$ worldsheet
is uniquely fixed by the chirality of the Wilson loop
and the connectivity of the structure.
Since there are no closed loops in the tube structure,
no ambiguity can result from this prescription.
Thus, to each $\son/\spn$ map, there corresponds a well-defined
$\sun$ map with the same weight
(since the twist point tubes have the same sign and
weight for $\sun$ and $\son/\spn$).

To demonstrate the isomorphism in the other direction,
take an $\sun$ worldsheet
and erase the chiralities of each of the components
to obtain the corresponding $\son/\spn$ worldsheet.

\REF\rTay{W. Taylor,  hep-th/9404175.}
It might be thought
that we have neglected a whole set of maps
in the $\sun$ string theory,
namely those containing infinitesimal area-dependent tubes.
(There are no area-dependent tubes in $\un$.)
In fact,
to leading order in $1/N$,
the contributions of area-dependent OPTs and ORTs
(tubes that connect sheets of the same or opposite
chirality respectively)
to the Wilson loop VEV exactly cancel [\rTay].
To see this,
consider some worldsheet connected by an OPT
to an otherwise disconnected vacuum piece $V$.
(This is the only type of worldsheet allowed in leading order.)
Then consider the same worldsheet connected by an ORT
to $\overline{V}$,
where $\overline{V}$ is obtained from $V$ by reversing the chiralities
of each of its components.
The weights of these two maps are equal
up to an overall sign,
and so they cancel.
Thus, only (area-independent) twist-point tubes
contribute in $\sun$ to leading order in $1/N$.

By exhibiting the large $N$ isomorphism of the string maps
in the $\sun$, $\un$, $\son$, and $\spn$ string theories,
we have just demonstrated the universality
of normalized Wilson loops $\nW_{(n)}$
for conjugacy classes with a single cycle on $S^2$.
Next, we consider a normalized Wilson loop VEV $\nWk$,
where the conjugacy class $\kappa$ has more than one cycle,
$\kappa = (n_1)(n_2)\ldots (n_c)$.
To leading order in $1/N$,
it is clear that the maps
that contribute to this Wilson loop expectation value
will simply be {\it disconnected} products of the
maps that contribute to each of the cycles $(n_i)$.
(Dividing by $Z$ removes disconnected
{\it vacuum} components for $\Wk$,
but the maps may still be disconnected.)
Any connected map will be suppressed by at least one power of $1/N$.
Thus,
to leading order in $1/N$,
$$
\nWk = \prod_{i=1}^{c}  \nW_{(n_i)} \largeNlimit  O \left(N^c\right)
\eqn\eProduct
$$
so from the previous result,
it follows that $\nWk$ is identical for each of the groups
in the large $N$ limit.
The same argument shows that the expectation value
of an arbitrary number of non-intersecting homologically trivial
Wilson loops on a sphere is universal.
This is because the expectation value
is the product of the expectation values for each
of the separate Wilson loops.

The proof is easily extended to homologically trivial Wilson loops
on the torus and on higher genus orientable target spaces.
In these cases, the only string maps that contribute to leading
order in $1/N$ are those with a single component $C_0$ with
the topology of the disk;
maps with any twist-point tubes are easily seen to be sub-leading.
This greatly reduced set of contributing worldsheets leads
to much simpler expressions for the universal VEVs
for homologically trivial Wilson loops
on any orientable surface of genus $h \ge 1$; for example,
$$
\eqalign{
\nW_{(1)}  & \largeNlimit N \e^{-\half A_1 },   \cr
\nW_{(2)}  & \largeNlimit N (1-A_1) \e^{- A_1 },   \cr
\nW_{(3)}  & \largeNlimit N \left( 1-3A_1 +\threehalf A_1^2 \right)
\e^{-\threehalf A_1 }.   \cr
}\qquad h \ge 1
\eqn\eWhighergenus
$$

Wilson loops also have universal leading-order
expectation values on nonorientable surfaces.
For the homologically trivial Wilson loop $\nW_{(1)}$ on $RP^2$,
the worldsheets that contribute to leading order in $1/N$
will have the same structure as those for $S^2$.
Each worldsheet will consist of connected components
$C_0$, $C_1, \cdots, C_m$,
where $C_0$ has the topology of a disk
and the others that of a sphere.
The maps from $C_i$ to $RP^2$ must all have even degree.
The $\son$ and $\spn$ string theories
also allow maps of odd degree,
corresponding to nonorientable worldsheets,
but these will all be higher order in $1/N$
relative to the orientable worldsheet contributions
since they involve cross-cap insertions.
Thus, there is again an isomorphism of maps, and an
equality of Wilson loop expectation values.
For example,
the normalized Wilson loop
in the fundamental representation
has the universal leading order expectation value
$$
\nW_{(1)}   \largeNlimit
 N \bigg[ \e^{-\half A_1 }  +  \e^{-\half A_1 -  A_2}
 +  \big( \half A_1^2 - A_1 - 1 \big)
\e^{-\threehalf A_1 - A_2} + \cdots \bigg]  .
\eqn\eWRP
$$
For nonorientable surfaces of higher genus,
the only string maps that contribute in leading order
in $1/N$ to homologically trivial Wilson loops
are those with a single component $C_0$ with
the topology of the disk.

Homologically non-trivial Wilson loops
are also universal in the large $N$ limit.
For example,
the homologically non-trivial Wilson loop on the torus
with $\kappa = (1, \bar{1})$
(which reduces to $\kappa = (1^2)$ for $\son/\spn$)
has the universal large $N$ normalized VEV
$$
\nW_{(1,\bar{1})}
\largeNlimit
2 \left( \e^{-\half A }   + \e^{-A}
+ \e^{{3 \over 2} A} + \e^{-2A} + \cdots \right)
= { 2\over {\e^{\half A} -1}},
\eqn\eNontriv
$$
where $A$ is the dimensionless area of the torus.
The terms in the sum correspond to
multiple covers of the cylinder by the cylinder.
We anticipate no obstacles to extending the string proof
of universality to the general case of
homologically non-trivial Wilson loops
on higher genus surfaces.

%
\chapter{Phase transitions}

Large $N$ Yang-Mills theories on the two-sphere
have a strong- and weak-coupling phase
separated by a third-order phase transition [\rDK, \rCS].
Although the most direct way to understand this
is in terms of the related matrix model,
it is interesting to examine
this phase transition
in string-theoretic terms [\rTay].
Toward this end,
we call attention to the
behavior of the string sums
contributing to Wilson loops
as one approaches the critical point.

Restricting our attention to $\sun$ Yang-Mills theory
in this section,
we recall that the string theory of QCD$_2$
is formulated as a strong-coupling expansion
that splits into two nearly independent chiral sectors [\rGTM].
(This split, and the individual chiral theories themselves,
may not be sensible for all coupling.)
We can test the convergence
of the contributions to the
individual chiral sectors on $S^2$
by studying the expansion for the Wilson loop directly,
analogous to the calculation carried out by Taylor
for the partition function [\rTay].

We may pick out the contributions
to the Wilson loop VEV in the fundamental representation
on the sphere \eWone~due to maps
from a single chiral sector
containing only mobile branch points.
This subset of string maps
corresponds to the ``${\rm B1}$'' string theory of ref.~[\rTay].
In the large $N$ limit,
only simply-connected maps in the ${\rm B1}$ theory contribute,
and for a single chiral sector,
these maps possess mobile branch points in area $A_1$ only
(cf. eq.~\eWone)
$$
\eqalign{
{ \nW_{(1)} (A_1, A_2)  \over N}
& \largeNlimit \e^{-\half A_1 }
+  \half A_1^2  \e^{-A_1 - \half A_2}
+  \half A_1^4  \e^{-\threehalf A_1 - A_2} + \cdots
\cr
& = \sum_{n=1}^\infty \Gamma_n {{A_1^{2n-2}}\over{(2n-2)!}}
\e^{-\half n A_1 - \half (n-1) A_2}.
\cr}
\eqn\eB
$$
The coefficient $\Gamma_n$
counts the number of distinct string maps
from surfaces consisting of $n-1$ spheres and 1 disk
simply-connected by $2n-2$ branch points
to the target space disk.
A closely related combinatorial problem
[\rTay, \rCT]
is the number of distinct string maps
from $n$ spheres simply-connected by $2n-2$  branch points
to the sphere
$$
G_n = n^{n-3}{{(2n-2)!}\over{n!}}.
\eqn\eGn
$$
These coefficients are related by $\Gamma_n = n G_n$,
since the disk is distinct from the rest of the worldsheet
(spheres)
and thus the overall symmetry factor is reduced.

Equation \eB~converges for $A_1$ sufficiently large,
regardless of the value of $A_2$.
Taking $A_2$ to zero in \eB,
the resulting expression is convergent
for $A_1$  greater than a certain
area $A_{c}$,
$$
{ \nW_{(1)} (A_1, 0)  \over N}
\largeNlimit \sum_{n=1}^{\infty} {{n^{n-2}}\over{n!}}A_1^{2n-2}
\e^{-\half nA_1} .
\eqn\eXral
$$
This expression has identical convergence properties
to that of the ${\rm B1}$ free energy on a sphere of area $A_1$,
and converges outside the region
$0.73 \lsim A_1 \lsim 11.9$.
This divergence below $A_1 \approx 11.9$
may be related to the
well-known third-order phase transition.
The relation between the non-analytic behavior
of quantities such as the free energy
and that of Wilson loop VEVs
is poorly understood.
Above we have shown that the
behavior of {\it part} of the
Wilson loop VEV in two dimensions
mimics the behavior of the free energy at large $N$.

%
\chapter{Conclusions}

Wilson loops on an arbitrary surface
have universal large $N$ VEVs for all gauge groups.
This fact is unexpected and obscure in
the gauge theory expressions for these observables,
but is easily understood from the underlying string description.
In this paper, we have used the string description
to present a general proof of
the large $N$ universality of Wilson loops.
Understanding universality
in the strong-coupling limit
is thus a significant application of the string approach.

The analysis and results of this paper
complement those of ref.~[\rCS],
in which universality was first
found using large $N$ matrix model techniques.
Furthermore, we have also shown that the chiral part of the
large $N$ expression
for an $\sun$  Wilson loop on the sphere
fails to converge below a certain coupling,
perhaps indicating the presence of a phase transition
as suggested in [\rTay].

\REF\rUniv{P. Horava, hep-th/9311156; \nextline
 S. Cordes, G. Moore, and S. Ramgoolam, hep-th/9402107.}
\REF\rGM{ D. Gross and A. Matytsin, \NPB 429 \rm (1994) 50;
hep-th/9410054.}
\REF\rCNS{M. Crescimanno, S. Naculich, and H. Schnitzer,
in preparation.}

These results suggest the existence of an
underlying universal theory on the worldsheet
to leading order in $1/N$,
such as that proposed in ref.~[\rUniv].
The $O(1/N)$ corrections
to these theories are not universal [\rNRS, \rRam].
Corrections to the $\sun$ partition function
in the weak coupling phase
have been worked out in ref.~[\rGM];
further developments will be presented in ref.~[\rCNS].

\bigskip
\noindent {\bf Acknowledgements}
\medskip
We thank K. Bardakci, H. Riggs, and W. Taylor for comments.
\vfill
\eject

\refout
\end